\documentstyle{article}
\newcommand{\bnabla}{\mbox{\boldmath$\nabla$}}
\begin{document}

\begin{center}

{\Large {\bf Schiff Theorem Revisited}}\\[0.3cm]

R.A. Sen'kov$^{1,2}$, N. Auerbach$^{3}$, V.V. Flambaum$^{4}$,
and V.G. Zelevinsky$^{1}$\\[0.3cm]


{\sl $^{1}$National Superconducting Cyclotron Laboratory and\\
Department of Physics and Astronomy, Michigan State University,\\
East Lansing, MI 48824-1321, USA\\
$^2$ Department of Physics, Novosibirsk State University,\\
Novosibirsk 630090, Russia \\
$^3$ School of Physics and Astronomy, Tel Aviv University, Tel
Aviv, 69978, Israel \\
$^{4}$School of Physics, University of New South Wales, Sydney
2052, Australia }\\

\end{center}


\begin{abstract}
We carefully rederive the Schiff theorem and prove that the usual
expression of the Schiff moment operator is correct and should be
applied for calculations of atomic electric dipole moments. The
recently discussed corrections to the definition of the Schiff
moment are absent.
\end{abstract}



The search for interactions violating time reversal (${\cal T}$-)
invariance is an important part of studies of fundamental symmetries
in nature. The main hopes for the extraction of nucleon-nucleon and
quark-quark interactions violating fundamental symmetries emerge
from the experiments with atoms and atomic nuclei, see the recent
review \cite{ginges04} and references therein. The best limits on
${\cal P},{\cal T}$-odd forces have been obtained from the
measurements of the atomic electric dipole moment (EDM) in the
$^{199}$Hg \cite{rgf01} and $^{129}$Xe \cite{jacobs95} nuclei.


The hadronic part of the atomic dipole moment associated with the
EDM of the nucleus manifests itself through the Schiff moment which
is the first non-vanishing term in the expansion of the nuclear
electromagnetic potential after including the screening of the
atomic electrons \cite{pr50,sch63,san67}. The standard expression
for the operator of the Schiff moment was repeatedly derived, see
for example in \cite{fks84},
\begin{equation}
S_k = \frac{1}{10}  \int \left( x_{}^2 x_k - \frac{5}{3}
\left\langle
   x_{}^2 \right\rangle_{\rm ch} x_k - \frac{2}{3}  \langle Q_{k k'}
\rangle x_{k'} \right) \rho ({\bf x}) d^3 x.   \label{1}
\end{equation}
Here $k,k'$ are the Cartesian vector components, $\rho({\bf x})$ is
the ground state nuclear density, $\langle x^{2}\rangle_{{\rm ch}}$
and $\langle Q_{kk'}\rangle$ are the nuclear charge mean square
radius and the expectation value of the quadrupole tensor,
respectively. Note that the nuclei of current experimental interest,
$^{199}$Hg, $^{129}$Xe, $^{225}$Ra, have nuclear spin $I=1/2$, so
that $\langle Q_{k k'} \rangle=0$ (spin $I=1/2$ provides certain
experimental advantages since the levels $I_{z}=\pm 1/2$ are not
split by external electric fields and have small collisional
broadening).

The exact form of the Schiff moment operator is important for the
correct interpretation and analysis of experimental data, evaluation
of future experimental plans with the best nuclear candidates and
for the search for possible corrections; the current status of
experimental efforts and corresponding theoretical discussions can
be found on the website of the INT workshop \cite{workshop}. The
Schiff theorem and the result (\ref{1}) for the Schiff moment were
put in doubt in the recent paper \cite{haxt}. The goal of our paper
is to rederive the conventional form of the Schiff moment. To make
our discussion as simple and transparent as possible, we present the
consistent derivation of the Schiff moment operator and make the
comparison with the derivation in Ref. \cite{haxt}.

Let us consider a neutral atom in a uniform external electric field
${\bf E}_{\rm ext}$. Neglecting the magnetic interaction, the
Hamiltonian of the entire system can be written as
\begin{equation}
H_{\rm Atom} = H_{\rm Electrons} +
   H_{\rm Nucleus} + \sum_{i=1}^Z \left( e \Phi ({\bf r}_i) - e
   {\bf r}_i \cdot{\bf E}_{\rm ext} \right) - {\bf d}_N\cdot{\bf E}_{\rm ext},
                                                             \label{2}
\end{equation}
where ${\bf r}_{i}$ label the electron coordinates, while ${\bf
d}_N$ is the operator of the nuclear electric dipole moment. The
nuclear electrostatic potential $\Phi ({\bf r})$ can be expressed
through the nuclear charge density $\rho ({\bf x})$,
\begin{equation}
\Phi( {\bf r})=\int\frac{\rho( {\bf x})\,d^3 x}{|{\bf x} - {\bf
r}|}.                                          \label{3}
\end{equation}

A conventional way to derive the Schiff moment is to make a unitary
transformation $e^{i U}$ with a suitable Hermitian operator $U$ that
will be chosen in the form (see for example, Appendix to Ref.
\cite{SAF97})
\begin{equation}
U = \frac{\langle {\bf d}_N\rangle}{Z | e |}\cdot \sum_{i = 1}^Z
{\bf p}_i.                        \label{4}
\end{equation}
Here ${\bf p}_i$ are the momentum operators of atomic electrons and
$\langle {\bf d}_N\rangle$ denotes the expectation value (allowed
due to ${\cal P}$- and ${\cal T}$-violation) of the nuclear electric
dipole moment with the exact nuclear ground state wave function. We
have to stress that the mean dipole moment introduced here is a
$c$-number that is small, being induced by the weak interactions.
Then the result of the unitary transformation of the Hamiltonian
(\ref{2}) can be written as
\begin{equation}
H'_{\rm Atom} = e^{i U}H_{\rm Atom} e^{- i U} \approx H_{\rm Atom} +
i [ U, H_{\rm Atom}].                       \label{6}
\end{equation}
The commutator
\begin{equation}
 i [U, H_{\rm Atom}] = \langle {\bf d}_N \rangle
 \cdot\left({\bf E}_{\rm ext} - \frac{1}{Z}  \sum_{i = 1}^Z  {\bnabla}_i
   \Phi ({\bf r}_i) \right) = \left\langle {\bf d}_N \right\rangle\cdot  \left(
   {\bf E}_{\rm ext} +{\bf E} (0) \right)       \label{7}
\end{equation}
has a clear physical meaning. Indeed, the second term in the
parentheses can be interpreted as the average electric field $ {\bf
E} (0)$ produced by atomic electrons and acting on the nucleus. For
an exact commutator with the Hamiltonian, the average value in a
stationary state of discrete spectrum vanishes, $\langle
\Psi|[U,H]|\Psi\rangle=0$. It means that the total electric field
acting on the nucleus equals to zero, and the nuclear electric
dipole moment is screened (Schiff theorem).

The unitary transformation changes the Hamiltonian (\ref{2}) by
adding the two terms in (\ref{7}). The first term transforms the
interaction of the nuclear dipole moment with external electric
field to
\begin{equation}
- \Bigl( {\bf d}_N - \langle {\bf d}_N
  \rangle \Bigr)\cdot  {\bf E}_{\rm ext}.     \label{8}
\end{equation}
The electric dipole moment is defined as a variational derivative of
energy with respect to the weak external electric field so that we
are interested only in the first-order correction to atomic energy
that vanishes as the {\sl expectation value} of the difference,
$({\bf d}_N - \langle {\bf d}_N\rangle)$, in Eq. (\ref{8}). Because
of the cancellation, the expression (\ref{8}) does not contribute in
the first order to the energy shift of the ground state and
therefore to the dipole moment of the atomic system as well. It
still will contribute to other observables, such as nuclear
polarizability, in the higher orders of perturbation theory.

The remaining second term of Eq. (\ref{7}) changes the interaction
of atomic electrons with the nucleus. Instead of the usual
electrostatic potential, we should consider
\begin{equation}
 \Phi ( {\bf r}) - \frac{1}{Z e} \langle
  {\bf d}_N \rangle \cdot{\bnabla} \Phi ( {\bf r}). \label{9}
\end{equation}
This is the basic expression for derivation of the nuclear Schiff
moment. Let us emphasize that for calculating the atomic EDM we need
to know only the expectation value of Eq. (\ref{9}) for the nuclear
ground state wave function. Indeed, after cancellation of the
nuclear EDM, see (\ref{8}), the atomic EDM can be written as
\begin{equation}
{\bf d}_{\rm Atom} = \sum_n\frac{\langle 0| e\sum_i^Z {\bf r}_i
|n\rangle\langle n|e\sum_i^Z \left( \Phi ( {\bf r}_i) - \frac{1}{Z
e} \left\langle {\bf d}_N \right\rangle\cdot {\bnabla} \Phi (
{\bf r}_i)\right)|0\rangle}{E_0-E_n} + {\rm h.c.}  \label{10}
\end{equation}
We assume here the factorization of the nuclear and electronic wave
functions (see the short discussion of corrections below).

The operator of the nuclear Schiff moment can be now derived in
various ways. We will expand the nuclear charge density in gradients
of the delta-function, the method often used in effective theories:
\begin{equation}
\rho ({\bf x}) = a \delta ( {\bf x}) + b_k \nabla_k \delta ( {\bf
x}) + \frac{c_{k k'}}{2} \nabla_k \nabla_{k'} \delta ( {\bf x}) +
\ldots.                                   \label{11}
\end{equation}
The coefficients of the expansion can be found according to
\begin{equation}
  \int \rho ( {\bf x}) d^3 x = Z|e|,        \label{12}
\end{equation}
\begin{equation}
\int {\bf x} \rho ( {\bf x}) d^3 x = \left\langle {\bf d}_N
\right\rangle,                               \label{13}
\end{equation}
\begin{equation}
\int x^2 \rho ({\bf x}) d^3 x = Z|e| \left\langle x^2
  \right\rangle_{\rm ch},                    \label{14}
\end{equation}
\begin{equation}
\int  \left( 3 x_k x_{k'} - \delta_{k k'} x^2 \right) \rho ({\bf x})
d^3 x= Z|e| \left\langle Q_{k k'} \right\rangle, \label{15}
\end{equation}
\begin{equation}
\int x^2  {\bf x} \rho ( {\bf x}) d^3 x = \left\langle {\bf O}
\right\rangle.                          \label{16}
\end{equation}
Finally we come to the following density expansion:
\[\rho ( {\bf x}) = \rho^{(0)} ( {\bf x}) + \rho^{(1)} ({\bf x}) + \rho^{(2)}
  ({\bf x}) + \ldots\]
 \[ = Z|e| \left[ \delta ({\bf x}) + \frac{\left\langle x_{}^2
  \right\rangle_{\rm ch}}{6} \bnabla^{2} \delta ({\bf x}) + \ldots \right]\]
\[  -\left[ \left\langle {\bf d}_N \right\rangle\cdot {\bnabla}
\delta ({\bf x}) + \frac{\left\langle {\bf O} \right\rangle\cdot
{\bnabla} }{10} \bnabla^{2} \delta ({\bf x}) + \ldots \right]\]
\begin{equation}
  + Z|e| \left[ \frac{\left\langle Q_{k k'} \right\rangle}{6} \nabla_k
  \nabla_{k'} \delta ({\bf x}) + \ldots \right] + \ldots \label{17}
\end{equation}
For the Schiff moment we need to consider only the dipole part of
this expansion in the first term of Eq. (\ref{9}), $\Phi ({\bf r})$,
and the monopole part for the second term of Eq. (\ref{9}), $ (1/Z
e)\langle{\bf d}_N\rangle\cdot {\bnabla}\Phi({\bf r})$. This leads
to the conventional form (\ref{1}) for the nuclear Schiff moment.

The expectation value of the expression (\ref{9}) in the nuclear
ground state $|0_{N}\rangle$ is given by
\begin{equation}
\langle 0_{N}| e \Phi ({\bf r}) - \frac{1}{Z}\langle
  {\bf d}_N\rangle\cdot {\bnabla} \Phi ({\bf r})|0_{N}
  \rangle = - \frac{Z e^2}{| {\bf r} |} + 4 \pi e {\bf S}\cdot {\bnabla}
  \delta ({\bf r}) + \ldots,               \label{18}
\end{equation}
where ${\bf S}$ is the expectation value of the nuclear Schiff
moment (\ref{1}).

The derivations in Ref. \cite{haxt} and presented here are quite
similar. The authors of \cite{haxt} used a commutator, analogous to
that we had in Eq. (\ref{7}), in a little different way with the
generator $U$ equal to
\begin{equation}
 U = \frac{ {\bf d}_N }{Z | e |}\cdot  \sum_{i =
   1}^Z {\bf p}_i ,                   \label{19}
\end{equation}
where $ {\bf d}_N $ is the operator of the nuclear dipole moment,
rather than a $c$-number. After that the cancellation we have
observed in Eq. (\ref{8}) occurs not only for the ground state
expectation value but for all matrix elements identically,
including, for example, the calculation of the usual nuclear
polarizability that does not require any symmetry violation. But
treated as an operator, $ {\bf d}_N $ does not commute with the
nuclear part of the Hamiltonian, and this is the source of further
corrections. One cannot treat the expectation value of the operator
product $({\bf d}_{N}\cdot \bnabla \Phi)$ as a product of the
expectation values; each of the factors has large matrix elements to
the excited nuclear states and, considered in an exact fashion,
brings two-body correlations in the Schiff moment operator which are
essentially an artefact of the approach used in Ref. \cite{haxt}.


Summarizing, we believe that the usual way of derivation of the
Schiff moment operator and the resulting form (\ref{1}) of the
Schiff moment operator itself are correct. New contributions to the
Schiff moment associated with two-body correlations in reality do
not appear. The way suggested in \cite{haxt} is more complicated and
should be treated with high accuracy. In addition we can mention
that the term in the Schiff moment (\ref{1}) that contains the
ground state expectation value of the nuclear quadrupole tensor also
emerges in the standard derivation. This term is usually neglected
in practical calculations; even for strongly deformed nuclei with
spin $I>1/2$ the corresponding correction does not exceed 20\%.

Let us briefly discuss other corrections to the Schiff moment and
atomic EDM mentioned as important in Ref. \cite{haxt}. The largest
correction is due to the relativistic character of the electron wave
functions which vary inside the nucleus. This makes the
$\delta$-function expansion (\ref{11}) invalid. The generalized
theory was developed in Ref. \cite{FG02}. The more accurate
expression, the so-called local dipole moment, coincides with the
Schiff moment in the limit of $Z\alpha\ll 1$, and the corrections
start with the term $\propto (Z\alpha)^{2}$. The numerical
calculations were performed in \cite{DF05}.

It was explained in the pioneering paper by Schiff \cite{sch63} that
the screening theorem is violated by the hyperfine magnetic
interaction between the electrons and the nucleus. In fact this
effect gives a dominating contribution in light nuclei, hydrogen
\cite{sch63} and helium \cite{dzuba07}. However, in heavy nuclei of
experimental interest, the Schiff moment contribution is by orders
of magnitude greater. Indeed, the hyperfine interaction grows
$\propto Z$, while the contribution of the Schiff moment is $\propto
Z^{2}$ multiplied by a relativistic factor that is of the order of 1
for $Z=1$ and increases to 10 for $Z=80$ \cite{fks84}.

The assumption of the factorization of the atomic wave function into
a product of the nuclear and electron parts is not precise. The
corresponding correction produced by the virtual nuclear and
electron excitation was expressed in terms of nuclear polarizability
and evaluated in \cite{FGM07}; it is small for nuclei of current
experimental interest.

In addition, the contribution of the internal nucleon EDM was widely
discussed, see references in \cite{ginges04}. If the nucleons have
their own internal electric dipole moments, the form of the nuclear
Schiff moment should be extended. The total nuclear Schiff moment is
the sum of the usual part $S^{(0)}_k$ (see Eq. (1)) and the
contribution due to internal nucleon EDM ${\bf d}_a$.

\begin{equation}
S_k = S^{(0)}_k + \sum_a^{A} \left[ \frac{1}{6}  \left( x^2_a -
\langle
  x^2 \rangle_{{\rm ch}} \right) d_{a;k} + \frac{1}{15}  \left(
Q_{a;k
  k'} - \langle Q_{k k'} \rangle \right) d_{a; k'} \right].
\end{equation}
Experimentally, only the total Schiff moment can be observed.

This work was supported by the NSF grant PHY-0555366, Australian
Research Council, and the grant from the Binational Science
Foundation USA-Israel. R.A.~Sen'kov acknowledges support from the
Council of the President of the Russian Federation for the State
Support of Young Scientists (project: MK-2982.2006.2) and Dynasty
Foundation.

\end{document}